\documentclass[aps,reprint,pra,superscriptaddress,floats,floatfix,nobibnotes,longbibliography]{revtex4-2}

\usepackage{amsmath}
\usepackage{amsfonts}
\usepackage{amssymb}
\usepackage{graphicx}
\usepackage{color}
\usepackage[colorlinks=true,citecolor=blue,linkcolor=blue,urlcolor=blue]{hyperref}
\usepackage{times}
\usepackage{amstext}
\usepackage{latexsym}
\usepackage{bm}
\usepackage{mathtools}
\usepackage{lipsum}
\usepackage{orcidlink}
\usepackage{soul}

\providecommand{\moy}[1]{\langle #1 \rangle}
\DeclarePairedDelimiter\bra{\langle}{\rvert}
\DeclarePairedDelimiter\ket{\lvert}{\rangle}
\DeclarePairedDelimiterX\braket[2]{\langle}{\rangle}{#1 \delimsize\vert #2}
\renewcommand{\dag}{^{\dagger}}
\renewcommand{\exp}[1]{e^{#1}}

\begin{document}


\title{Quantum Resonator as a Directional Quantum Emitter}

\author{Luiz O. R. Solak\,\orcidlink{0000-0002-4760-3357}}
\email{solakluiz@estudante.ufscar.br}
\affiliation{Departamento de Física, Universidade Federal de São Carlos, 13565-905 São Carlos, São Paulo, Brazil}
\affiliation{CESQ/ISIS (UMR 7006), CNRS and Universit\'{e} de Strasbourg, 67000 Strasbourg, France}
\author{Bruno L. Vermes\,\orcidlink{0009-0008-4425-7488}}
\affiliation{Universidade Estadual Paulista (UNESP), Instituto de Geociências e Ciências Exatas, 13506-900 Rio Claro, São Paulo, Brazil}
\author{Antonio S. M. de Castro\,\orcidlink{0000-0002-1521-9342}}
\affiliation{Departamento de Física, Universidade Estadual de Ponta Grossa, 84030-900 Ponta Grossa, Paraná, Brazil}
\author{Celso J. Villas-Boas\,\orcidlink{0000-0001-5622-786X}}
\affiliation{Departamento de Física, Universidade Federal de São Carlos, 13565-905 São Carlos, São Paulo, Brazil}
\author{Daniel Z. Rossatto\,\orcidlink{0000-0001-9432-1603}}
\email{dz.rossatto@unesp.br}
\affiliation{Universidade Estadual Paulista (UNESP), Instituto de Ciências e Engenharia, 18409-010 Itapeva, São Paulo, Brazil}

\begin{abstract}
Single-photon sources are essential for testing fundamental physics and for the development of quantum technologies. In this work a single-photon source is investigated, based on a two-photon Jaynes-Cummings system, where the resonator works as the quantum emitter rather than the two-level system. This role reversal provides certain advantages, such as robustness against losses from the two-level system (e.g., dephasing), as it remains in its ground state throughout the entire dynamics. This provides higher efficiency, purity, and indistinguishability compared to sources based on the usual Jaynes-Cummings model under the same parameter conditions in both models. Another advantage of this system is the possibility of direct conversion of a coherent excitation pulse with one photon on average to a single-photon pulse with efficiency, purity, and indistinguishability above $90\%$. Since the entire excitation pulse is consumed in the generation of a single photon, the system also minimizes energy waste. The potential for implementing the two-photon JC model across different platforms expands the possibilities for controlled single-photon generation in applications in quantum information processing and computation.
\end{abstract}

\maketitle

\section{Introduction}

\label{s1} Efficient and controllable single-photon generation is an imperative task for the advancement of quantum technologies \cite{Lounis2005, Sangouard2012,reimer,thomas}, especially in the realm of quantum computing and communication \cite{Knill, kok,PhysRevLett.98.190504,Obrien,Wang2019}. Single-photon sources can be based on stochastic nonlinear processes, such as parametric down-conversion \cite{burnham,hong} and four-wave mixing \cite{fan,sharping}. In addition, they can be based on quantum emitters, which are effective two-level quantum systems that deterministically emit photons individually (photon antibunching) \cite{walls}, among them stand out semiconductor quantum dots \cite{Wang2019-2,Tomm2021}, Rydberg atoms \cite{ornelashuerta,shishuai}, color centers \cite{Andrini2024}, and trapped ions \cite{Keller2004,Higginbottom2016}. 

Since the photons are usually emitted through spontaneous emission, the direction in which they propagate is random. However, the properties of a single-photon source can be enhanced by coupling the quantum emitter to a cavity mode \cite{Lounis2005,Kuhn2010,Esmann2024}, which channels the emitted photons into a well-defined spatial mode (improves the efficiency of collecting them). It also improves the photon production rate, as the cavity enhances the spontaneous emission rate of the quantum emitter due to the Purcell effect \cite{Lounis2005}. 
Nevertheless, a common feature of this scheme is that single-photon generation relies on atomic transitions, which can potentially reduce efficiency due to atomic losses (e.g. photons emitted in a direction other than the cavity axis or atomic dephasing).

{Recently, alternative approaches \cite{Yan2018,PhysRevA.101.063824,PhysRevA.102.053713} have been explored beyond single-photon sources based on the Jaynes–Cummings (JC) interaction exploiting the single-photon blockade \cite{PhysRevA.100.063834}. Particularly,} a two-photon Jaynes-Cummings system \cite{Sukumar1981,PhysRevA.25.3206,Aliskenderov1987} has recently been considered a promising setup for producing single photons \cite{felicetti2,rossatoprl2019,zou2pjc,Li2024,zhou2025}. This system has been investigated for decades on a wide range of platforms, including optical \cite{gauthier} and microwave \cite{brune} cavities, trapped ions \cite{vogel,PhysRevLett.76.1796,felicetti}, quantum dots \cite{PhysRevB.81.035302, Singh2020}, and superconducting circuits \cite{neilinger,PRXQuantum.4.030326}. Remarkably, circuit-QED schemes that can implement the two-photon JC model within the strong-coupling regime have recently been proposed \cite{felicetti2,Bertet2005, PhysRevA.98.053859, w12t-92qg}, {with notable experimental progress being made toward realizing multiphoton processes \cite{chang2020}}.
Meanwhile, a thorough investigation of the key attributes (e.g., efficiency, purity, and indistinguishability \cite{Bienfang2023}) of a controllable, on-demand single-photon source is still lacking in the literature, particularly regarding its implementation as such a source — which is the aim of this study. In addition, we show that a two-photon JC system can implement a directional quantum emitter, using the resonator as the emitter instead of the two-level system. Thus, the quantum emitter itself naturally emits the single photon in a specific direction, eliminating photon transfer from the two-level system to the resonator, thereby enhancing robustness against dissipation from the two-level system since it remains in its ground state. This occurs because in this kind of interaction the presence of the two-level system introduces an anharmonicity to the energy spectrum of the resonator mode in a way that makes it an effective two-level system when in the strong-coupling regime, allowing the cavity to absorb and emit one photon at a time due to the single-photon blockade phenomenon \cite{felicetti2,rossatoprl2019,zou2pjc,Li2024,zhou2025}.

The present work theoretically explores a single-photon source based on a two-photon JC system, with a quantum resonator playing the role of a quantum emitter instead of a two-level system. For clarity and without loss of generality, we illustrate our proposal using a system consisting of a single atom coupled to a single mode of a two-sided optical cavity via the two-photon interaction. Nonetheless, it is important to stress that our approach is general and applicable to any quantum platform that allows the implementation of the two-photon JC model.
We investigate the on-demand generation of single photons, emitted from one side of the cavity by resonantly driving the other with coherent pulses, with the atom remaining in its ground state. We show that the cavity indeed behaves as an effective two-level system (quantum emitter) when the atom-cavity coupling strength surpasses the dissipation rates of the system (strong-coupling regime), and it is capable of producing single photons with high efficiency, purity, and indistinguishability (above $90\%$). Remarkably, we show the robustness of this single-photon source to atomic losses, ensuring higher efficiency, purity, and indistinguishability compared to sources based on the usual JC model under the same parameter conditions in both models.

This paper is structured as follows. Section \ref{s2} outlines the driven two-photon JC model and the main attributes of a single-photon source. In Section \ref{s3}, we present the main results and discussions, while we provide our conclusions in Section \ref{s5}.

\section{System}

\label{s2} 
\begin{figure*}[t]
\centering
    \includegraphics[width=17.8cm]{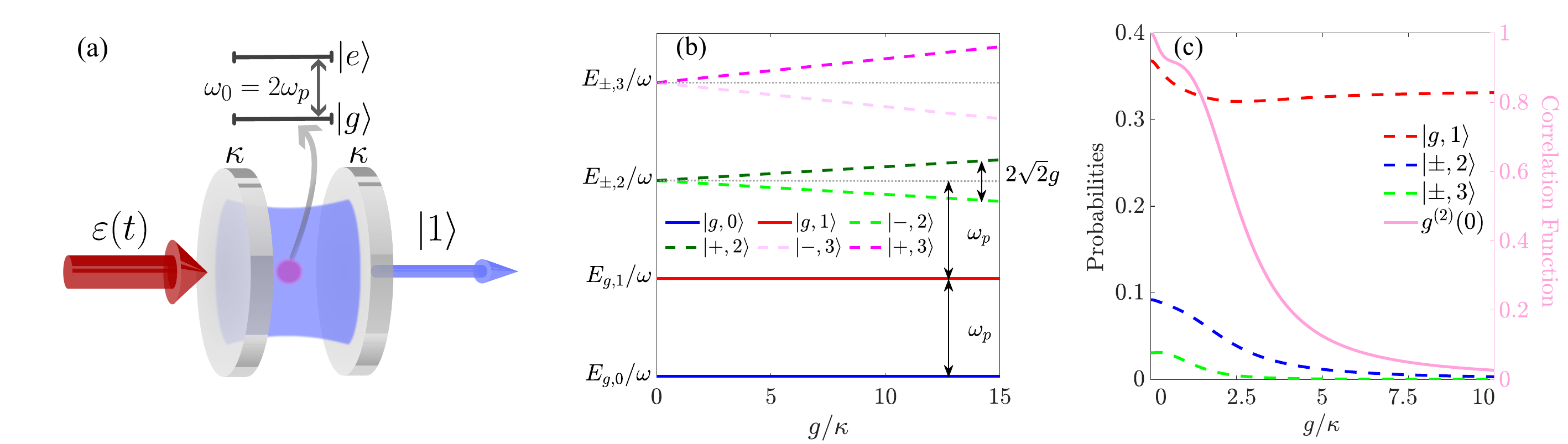}
    \caption{\label{sys_occup} (a) Pictorial illustration of the system setup. A two-level atom is confined in a two-sided cavity with decay rate of field amplitude $\kappa$. An incoming coherent pulse, with frequency $\omega_p$ and amplitude $\varepsilon(t)$, resonantly drives the cavity from one side, whereas a single-photon pulse is emitted at the other side. The atomic transition with frequency $\omega_0$ between the ground $\ket{g}$ and excited $\ket{e}$ states is resonantly coupled to the cavity mode through a two-photon interaction ($\omega_{0}=2\omega_{p}$). (b) Energy spectrum of the two-photon JC interaction when varying the coupling strength $g$. A two-photon transition is nonresonant with the first dressed states ($\ket{g,0}\rightarrow \ket{\pm,2}$) when the coupling $g$ is sufficiently strong, thus making the single-photon generation possible via the transition $\ket{g,0} \leftrightarrow \ket{g,1}$. In other words, the cavity becomes an effective two-level system, with the atom remaining in its ground state during the dynamics. (c) Eigenstates occupation probabilities and the second-order correlation function as functions of the coupling $g$ for a constant driving amplitude $\varepsilon(t)=\kappa$, showing that for strong enough coupling strength the system indeed have single-photon statistics, i.e., $g^{(2)}(0) \to 0$. Here we consider $\Gamma=\Gamma_{\phi}=\kappa/2$.}
    \label{fig:enter-label}
\end{figure*}

We consider a driven two-photon JC model \cite{felicetti2}, which is written in the interaction picture as ($\hbar = 1$)
\begin{equation} \label{HJC2}
    H = g(a^{2}\sigma_{+} + a^{\dagger 2}\sigma_{-}) + \varepsilon(t)(a\dag + a),
\end{equation}
in which a cavity mode with frequency $\omega$ is resonantly driven by a coherent field with frequency $\omega_p = \omega$ and modulated amplitude $\varepsilon(t)$, and is resonantly coupled (two-photon interaction) to a two-level atom with a transition frequency $\omega_0 = 2\omega$. Here, $a$ ($a\dag$) is the annihilation (creation) operator of the cavity mode, $\sigma_{-} = \ket{g}\bra{e}$ and $\sigma_{+} = \sigma_{-}^\dagger$ are the atomic ladder operators, with $\ket{g}$ standing for the atomic ground state while $\ket{e}$ for the excited state, and $g$ is the atom-cavity coupling strength. The system dynamics, including dissipative mechanisms under the white-noise limit \cite{qnoise}, is governed by the Lindblad master equation \cite{breuerpetruccione}
    \begin{align}\label{ME}
        \dot\rho = &-i\left[H,\rho\right]+\Gamma\left(2\sigma_{-}\rho\sigma_{+} - \sigma_{+}\sigma_{-}\rho - \rho\sigma_{+}\sigma_{-}\right) \nonumber\\ &+\kappa\left(2 a \rho a\dag - a\dag a \rho - \rho a\dag a\right) + \Gamma_{\phi}\left(\sigma_{z}\rho\sigma_{z} - \rho\right),    
    \end{align}
where $\Gamma$ and $\Gamma_{\phi}$ are the atomic polarization decay and dephasing rates, respectively, $\kappa$ is the decay rate of the cavity field amplitude, and $\sigma_z = \sigma_{+}\sigma_{-} - \sigma_{-}\sigma_{+}$. {This equation, together with all the results to be shown in this work, was numerically solved using QuTiP \cite{johansson,johansson2}.} The system setup is illustrated in \textbf{Figure} \ref{sys_occup}(a).

The eigenstates of the two-photon JC model are \cite{rossatoprl2019,zou2pjc}
    \begin{align}
        &\ket{g,n},\: \text{for}\: n<2, \\
        &\ket{\pm,n}=\frac{1}{\sqrt{2}}(\ket{g,n}\pm \ket{e,n-2}),\: \text{for}\: n\geq2,
    \end{align}
with respective eigenenergies
\begin{align}
    E_{g,n} &= \left(n-1\right)\omega ,\\
        E_{\pm,n}&=\left(n-1\right)\omega \pm g\sqrt{n(n-1)}\,.
\end{align}
Figure \ref{sys_occup}(b) shows that the subspace of the two lowest eigenstates is harmonic and uncorrelated (atom in the ground state), while higher-energy states form a JC doublet-like structure. For the system initially in its ground state $\ket{g,0}$ and ensuring that $g$ is strong enough, a resonant driving field is able to induce the system to the uncorrelated excited state only ($\ket{g,0}\to\ket{g,1}$), because the transition $\ket{g,1}\to\ket{\pm,2}$ (and so on) is not resonant with this driving field, thus preventing the entry of more than a single photon into the cavity mode (single-photon blockade) \cite{felicetti2,rossatoprl2019,zou2pjc,Li2024,zhou2025}. This behavior is illustrated in Figure \ref{sys_occup}(c) for a constant driving field amplitude [$\varepsilon(t) = \kappa$]. As $g$ increases, the only excited state that becomes populated is $\ket{g,1}$, while all other Fock components of the coherent driving field are reflected, except for the vacuum state. As a result, the cavity mode exhibits single-photon statistics, with the second-order correlation function $g^{(2)}(0)$, defined below, approaching zero.

Moreover, the strong coupling regime ($g\gg\kappa$) plays a major role for this model to be considered as a \textit{directional} quantum emitter. In this regime the system behaves effectively as a two-level system ($\ket{g,0}\leftrightarrow\ket{g,1}$) where the dynamics are independent from any {atomic excitation}. In this case one can rely on input-output theory, where for an empty two-sided cavity with identical mirrors, it is well stated that any resonant field impinging in one mirror will always leave the cavity from the other side \cite{walls,steck}, providing this model with directionality.

A single-photon source can be characterized by several figures of merit \cite{Bienfang2023}, including efficiency, purity, and indistinguishability \cite{thomas,PhysRevA.108.L031701}. Efficiency can be defined as the probability of detecting the emitted pulse in the single-photon state per trial (after each round of source excitation). Since a coherent pulse will be used as the excitation pulse, the efficiency can be computed as \cite{PhysRevA.108.L031701}
\begin{equation}
    \mathcal{E} = \kappa \textstyle\int \moy{a^{\dagger}(t)a(t)} dt .
\end{equation}
When the efficiency is not 100\%, the emitted pulse may have projection in the vacuum state or in multiphoton components. To ensure that only one photon is emitted per trial, the source must also exhibit high single-photon purity 
\begin{equation}
    \mathcal{P} = 1 - g^{(2)}{(0)},
\end{equation}
with $g^{(2)}{(\tau)}$ the normalized version of the second-order correlation function $G^{(2)}{(\tau)} = \int \moy{a^{\dagger}(t)a^{\dagger}(t+\tau)a(t+\tau)a(t)}dt$ in the Hanbury-Brown-Twiss experiment \cite{PhysRevA.108.L031701}. Lastly, indistinguishability, which is determined by the Hong-Ou-Mandel experiment, can be calculated as \cite{kaer}
    \begin{align}
        \mathcal{I} = \frac{\int^{\infty}_{0}dt\int^{\infty}_{0}d\tau|\langle a\dag(t+\tau)a(t)\rangle|^{2}}{\int^{\infty}_{0}dt\int^{\infty}_{0}d\tau \langle a\dag(t+\tau)a(t+\tau)\rangle\langle a\dag(t)a(t)\rangle} ,
    \end{align}
which certifies that the source produces identical photons when $\mathcal{I} \to 100\%$. 

\begin{figure*}[t]
\centering
    \includegraphics[width=17.8cm]{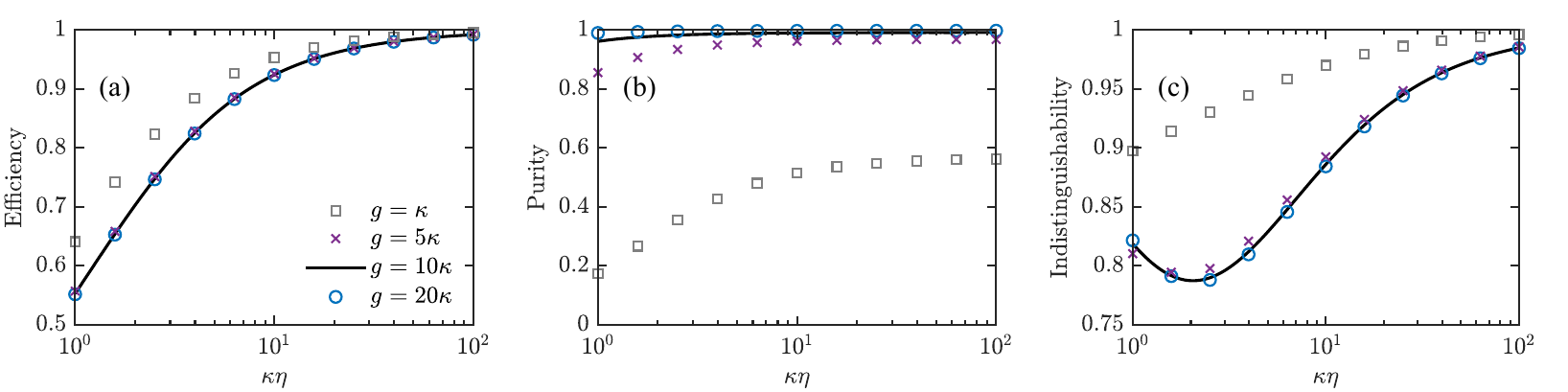}
    \caption{\label{fig:2} {(a)} Efficiency, {(b)} purity and {(c)} indistinguishability of the single-photon source as functions of the pulse duration $\eta$ for different values of $g$, considering ${\Omega}=\kappa$ and $\Gamma=\Gamma_{\phi}=0.5\kappa$.}
\end{figure*}

\section{Results} \label{s3}
To evaluate the performance of a two-photon JC system as a single-photon source, we consider{, for the sake of illustration,} a Gaussian coherent pulse as the excitation pulse, such that 
    \begin{align}
        \varepsilon(t) = \varepsilon_{\text{max}}\exp{{-}\frac{1}{2}\frac{(t-T)^{2}}{\eta^{2}}},
    \end{align}
with $\varepsilon_{\text{max}}={\Omega}/\sqrt{\eta\sqrt{\pi}}$, standard deviation $\eta$ (characteristic time of the pulse duration), and ${\Omega} = \sqrt{n_p}\kappa$, where $n_{p}$ is the mean number of photons in the input pulse. $T$ is the time at which the maximum amplitude of the pulse reaches the cavity. {It is important to stress that the following results are expected qualitatively for any pulse shape.}

For an excitation pulse with one photon on average (${\Omega}=\kappa \leftrightarrow n_p = 1$) and setting $\Gamma=\Gamma_{\phi}=0.5\kappa$, the efficiency, purity, and indistinguishability as functions of pulse duration $\eta$ for $1 \le g/\kappa \le 20$ are shown in \textbf{Figure} \ref{fig:2}. In this scenario, no substantial change in the behavior of the figures of merit is observed for $g \gtrsim 5\kappa$, especially when the pulse duration is sufficiently long ($\eta \gg \kappa^{-1}$). Moreover, for any $g$ in this range, the efficiency and indistinguishability approach 100\% as $\eta \gg \kappa^{-1}$, since this ensures that the spectral width of the excitation pulse fits the cavity linewidth. This prevents efficiency loss due to portions of the pulse being out of resonance with the cavity, which would otherwise be directly reflected by the cavity mirror. Then, a coherent pulse with one photon on average is effectively converted into a single-photon pulse, minimizing energy loss from the input pulse. The system achieves higher purity with increasing $g$. This is expected since the atom-field coupling is responsible for the anharmonicity in the energy spectrum of the two-photon JC model, which gives rise to the JC doublets [Figure \ref{sys_occup}(b)], preventing the excitation field from populating the dressed states and ensuring that no more than one photon enters the cavity at a time (single-photon blockade). Therefore, if ${\Omega}$ is increased, $g$ must also be increased accordingly to maintain the same degree of purity.  

\begin{figure}
\centering
    \includegraphics[width=8.5cm]{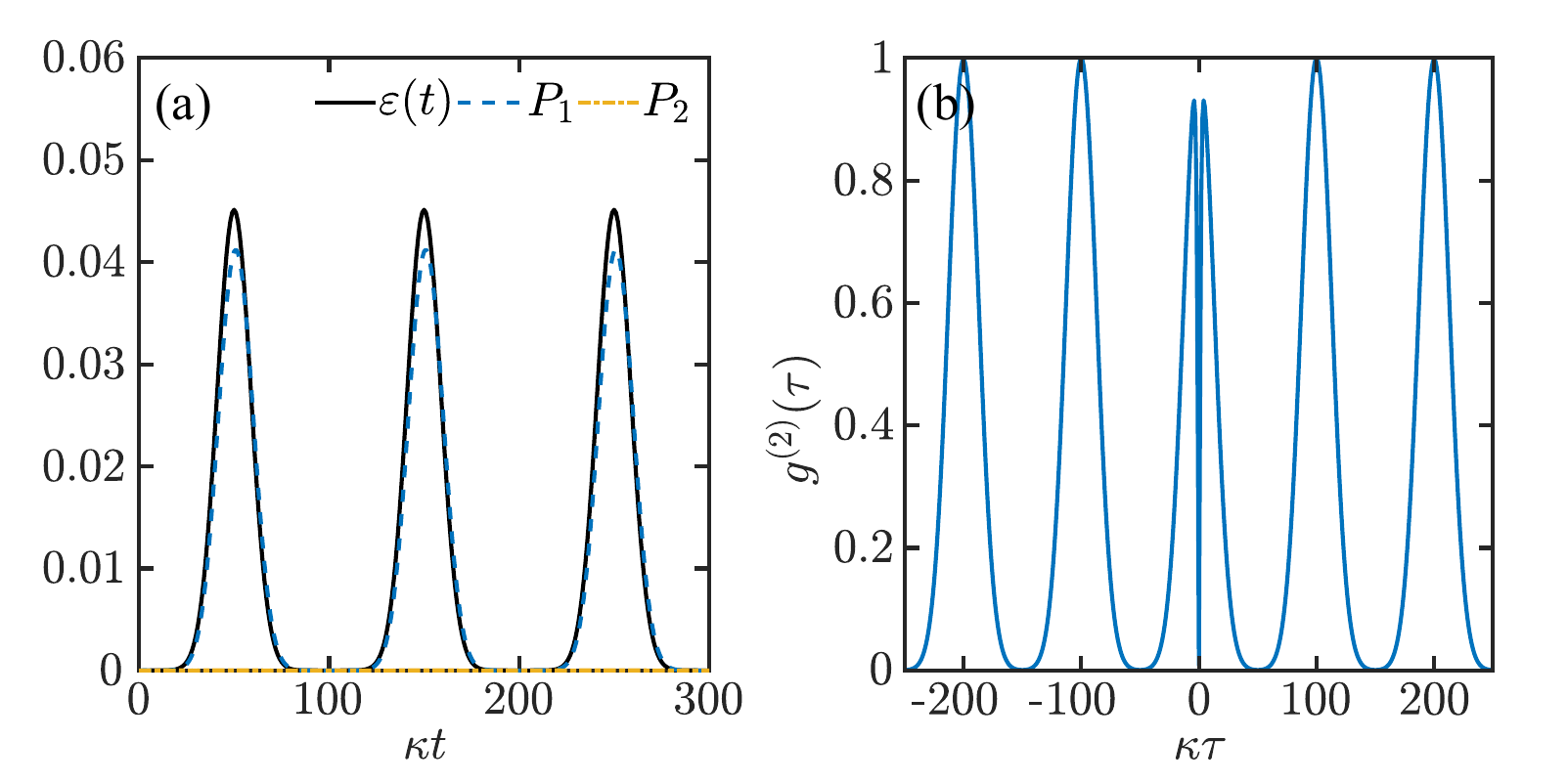}
    \caption{\label{fig:3} (a) Dynamics of the probabilities of finding the intracavity field in the Fock states $\ket{1}$ and $\ket{2}$ giving a pump pulse with modulated amplitude $\varepsilon(t)$, considering $g = 10\kappa$, ${\Omega} = \kappa$, $\eta = 12.5\kappa^{-1}$, and $\Gamma=\Gamma_{\phi}=0.5\kappa$. (b) Second-order correlation function $g^{(2)}(\tau)$ for the same parameters after a sequence of 500 pump pulses. In this scenario, the single-photon source has an efficiency of 94\%, a purity of 99\%, and an indistinguishability of 90\%.}
\end{figure}

Although $\{\mathcal{E}, \mathcal{P}, \mathcal{I}\} \to 1$ when $\eta \gg \kappa^{-1}$, it is important to note that a higher $\eta$ results in a slower single-photon generation rate, as more time is required per excitation cycle. Therefore, since a single-photon source must exhibit high values for all figures of merit, an optimal trade-off must be sought. As an example, in \textbf{Figure} \ref{fig:3}(a) we illustrate the dynamical response of the probabilities of finding the intracavity field in the Fock states $\ket{1}$ and $\ket{2}$ when the system is excited by $\varepsilon(t)$, considering $g = 10\kappa$, ${\Omega} = \kappa$, and $\eta = 12.5\kappa^{-1}$. {As expected from the eigenstates occupations, the remaining probability shown in this figure corresponds to the vacuum state, which is omitted for simplicity, since there is no probability of occupying higher excitation states in either the atom or the cavity field}. Figure \ref{fig:3}(b) shows the second-order correlation function after a sequence of 500 excitation pulses, using the same parameters. In this scenario, the system achieves high values for all figures of merit, namely $\mathcal{E} = 94\%$, $\mathcal{P} = 99\%$, and $\mathcal{I} = 90\%$.  {Furthermore, the system exhibits single-photon generation rate of $\kappa/100$, since in the pulsed excitation regime, the second-order correlation function exhibits a comb-like structure, with peaks separated by the pulse repetition period. Since each peak corresponds to correlations between photons emitted in successive excitation cycles, the temporal separation between adjacent peaks therefore gives the repetition period of the emission sequence \cite{Loudon2000}}.

\begin{figure*}
\centering
    \includegraphics[width=17.8cm]{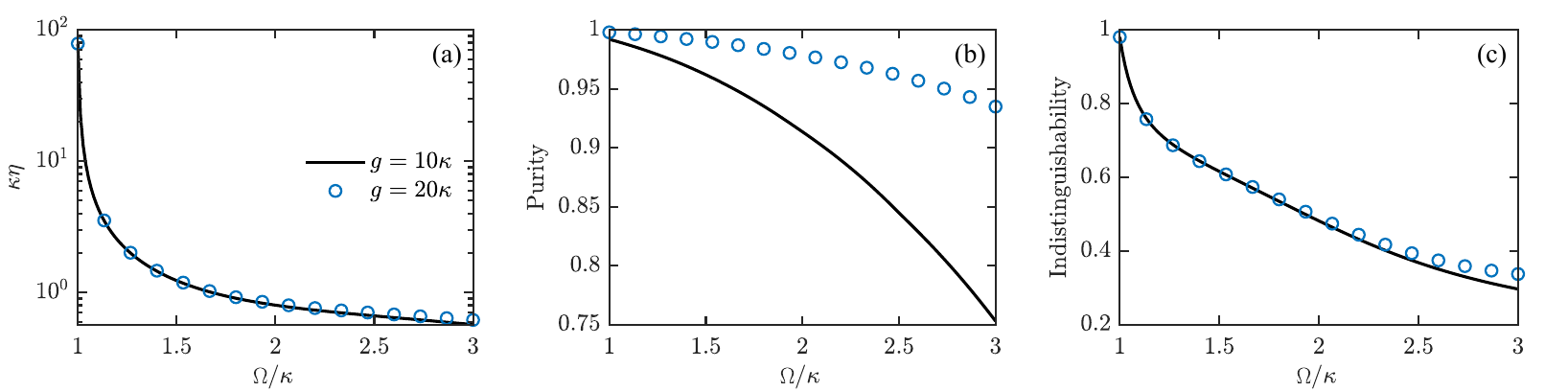}
    \caption{\label{fig:4} (a) Pairs of $({\Omega},\eta)$ that provides 99\% of efficiency for the single-photon source. Corresponding (b) purity and (c) indistinguishability as functions of $({\Omega},\eta)$ given in (a). Here we consider $\Gamma=\Gamma_{\phi}=0.5\kappa$.}
\end{figure*}

One way to increase the single-photon generation rate is to decrease $\eta$ while increasing ${\Omega}$, ensuring high efficiency and that only one photon enters the cavity per trial. However, in this case, a significant portion of the excitation pulse may be directly reflected, leading to wasted energy from the excitation pulse. In \textbf{Figure} \ref{fig:4}(a), we show pairs of (${\Omega},\eta$) for which $\mathcal{E} \approx 99\%$. As expected, Figure \ref{fig:4}(b) shows that, for a given value of $g$, purity decreases as ${\Omega}$ increases. As previously mentioned, $g$ must also be increased accordingly to maintain the same level of purity in this case. On the other hand, although the single-photon generation rate can be enhanced by increasing ${\Omega}$ while decreasing $\eta$, and the purity can still be maintained at a high level by sufficiently increasing $g$, this comes at the cost of reduced indistinguishability, as shown in Figure \ref{fig:4}(c). For this reason, it seems that ${\Omega} = \kappa$ presents a more favorable scenario for achieving high efficiency, purity, and indistinguishability while also avoiding energy waste, as practically the entire excitation pulse is absorbed by the cavity per trial.

\begin{figure}[t]
    \centering
    \includegraphics[width=8.5cm]{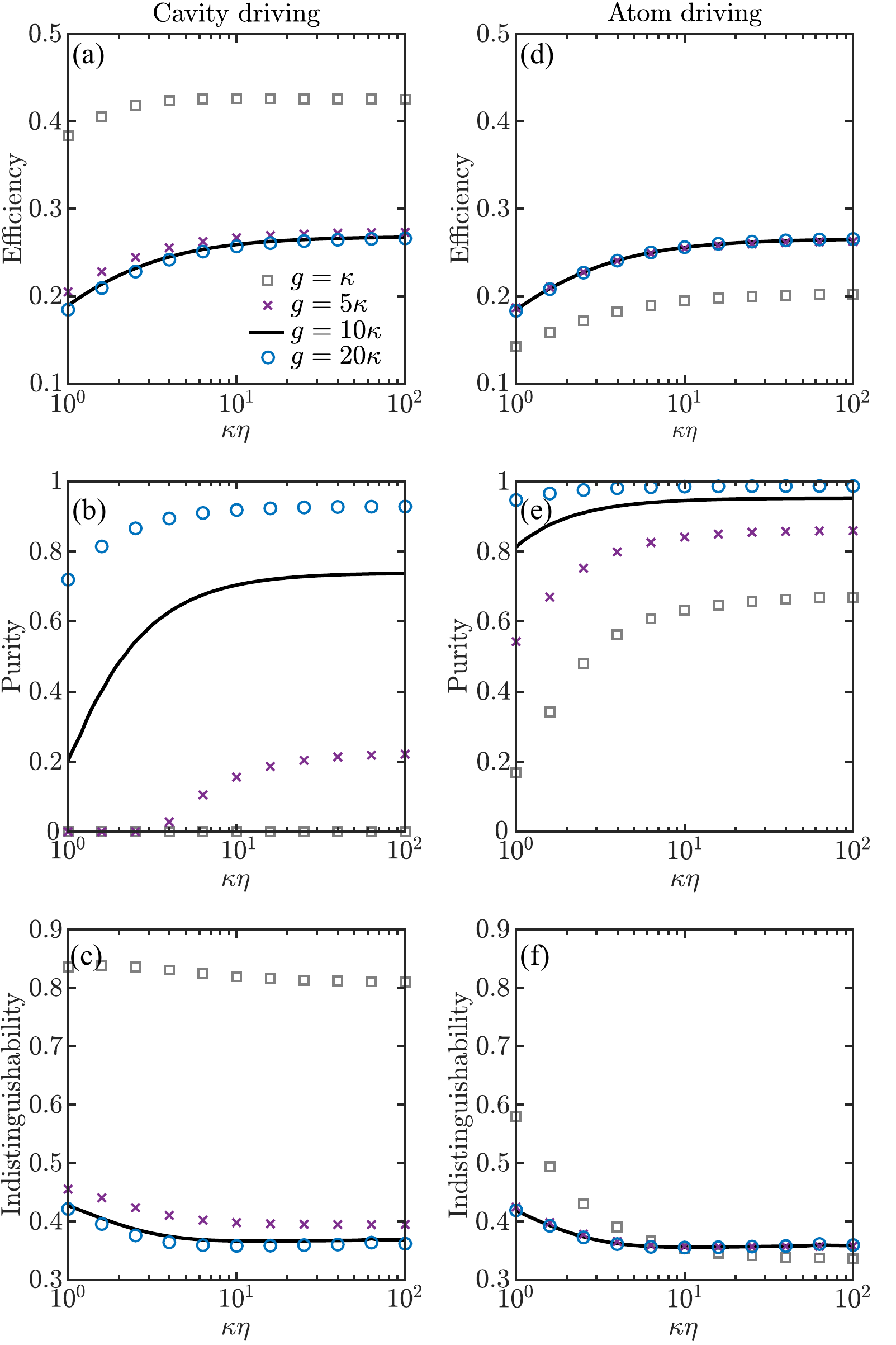}
    \caption{\label{fig:5} The efficiency, purity, and indistinguishability of a single-photon source based on the JC model are analyzed as functions of the pulse duration $\eta$ for different values of $g$, setting ${\Omega}=\kappa$ and $\Gamma=\Gamma_{\phi}=0.5\kappa$. We consider an excitation pulse driving either the cavity [{(a), (b) and (c)}] or the atom [{(d), (e) and (f)}].}
\end{figure}

Finally, it is important to emphasize the robustness against atomic dissipative processes of this implementation of a single-photon source based on the two-photon JC interaction, especially compared to the usual JC model employed in current sources based on quantum emitters \cite{Kuhn2010,Esmann2024}. For illustration purposes, the figures of merit considering the usual JC model (see Appendix), where the atom acts as the quantum emitter, are shown in \textbf{Figure} \ref{fig:5}, using the same parameters as in Figure \ref{fig:2}. We observe that a single-photon source based on the two-photon JC model exhibits better efficiency, purity, and indistinguishability than a source based on the usual JC model, the latter being stimulated either through the cavity (left panels in Figure \ref{fig:5}) or the atom (right panels in Figure \ref{fig:5}). For example, considering $g = 10\kappa$, ${\Omega} = \kappa$, $\eta = 12.5\kappa^{-1}$, and $\Gamma=\Gamma_{\phi}=0.5\kappa$ {(parameters lie well within the range achievable in current superconducting devices and in JC-based single-photon source implementations~\cite{Tomm2021})}, while the two-photon JC model yields $\mathcal{E} = 94\%$, $\mathcal{P} = 99\%$, and $\mathcal{I} = 90\%$, the usual JC model yields $\mathcal{E} = 26\%$, $\mathcal{P} = 71\%$, and $\mathcal{I} = 37\%$ for cavity driving and $\mathcal{E} = 26\%$, $\mathcal{P} = 95\%$, and $\mathcal{I} = 36\%$ for atom driving. This is because the usual JC model is not as robust to atomic dissipative processes as the two-photon JC model, since the JC system is stimulated through the transition $\ket{g,0} \to (\ket{g,1} \pm \ket{e,0})/\sqrt{2}$, with the dressed state having a component in the excited atomic state, making it susceptible to atomic dissipative processes. Therefore, in such a case, $\Gamma$ and $\Gamma_\phi$ must approach zero in order to optimize the figures of merit.

\section{Conclusion}
\label{s5} 
We have theoretically investigated the application of a two-photon JC system as a single-photon source. In this setup within the strong-coupling regime, the quantum resonator itself acts as a quantum emitter, leading to intrinsic directionality. Moreover, since the two-level system always remains in its ground state, this single-photon source is more robust against dissipative mechanisms from the two-level system than current sources based on the usual JC model, which can improve the efficiency, purity, and indistinguishability of the source. In particular, this system also offers the advantage of directly converting a coherent pulse with one photon on average into a single-photon pulse, thus preventing energy waste from the excitation pulse, as it is entirely consumed for the single-photon generation. Although we have employed the optical domain and atomic systems here to illustrate this type of single-photon source, an important practical aspect is the feasibility of implementing the two-photon JC model on different platforms {such as  optical \cite{gauthier} and microwave \cite{brune} cavities, trapped ions \cite{vogel,PhysRevLett.76.1796,felicetti}, quantum dots \cite{PhysRevB.81.035302, Singh2020}, and superconducting circuits \cite{neilinger,PRXQuantum.4.030326}}, especially in circuit-QED schemes that can enable the strong-coupling regime \cite{felicetti2,Bertet2005, PhysRevA.98.053859, w12t-92qg}, broadening the prospects for controlled single-photon generation in applications related to quantum information and computation.

\section*{Appendix: Single-photon Source Based on JC model} \label{app}

\begin{figure}
\centering
    \includegraphics[width=5.5cm]{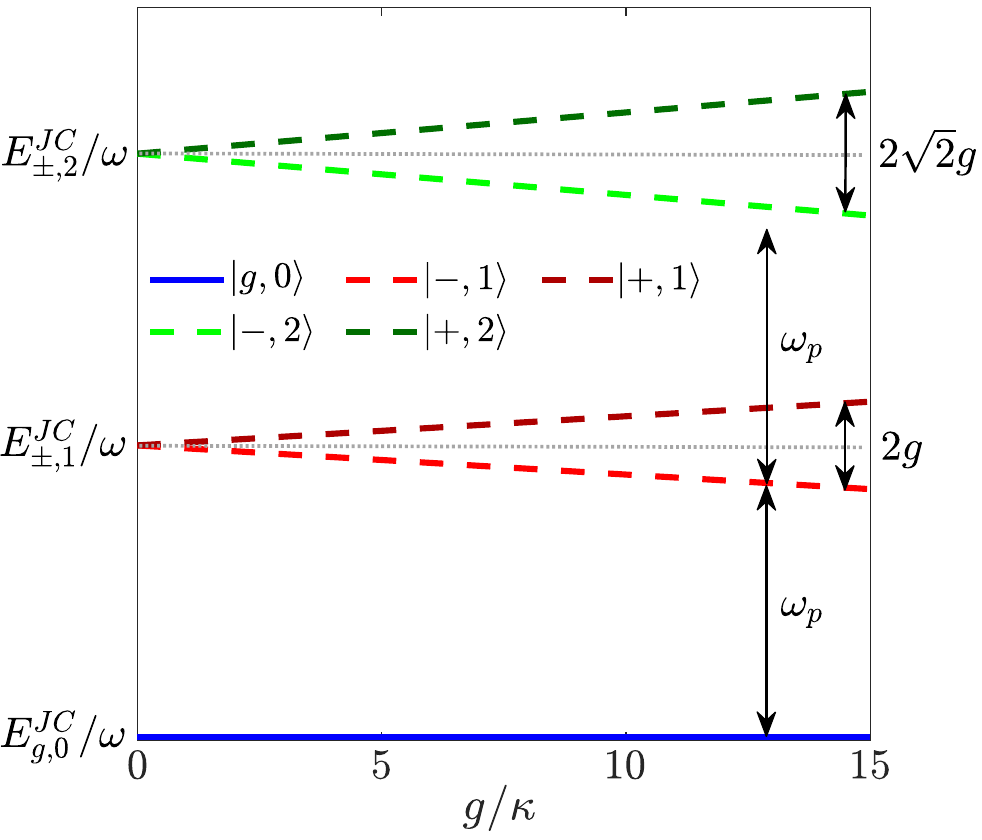}
    \caption{Energy spectrum of the JC interaction when varying the coupling strength $g$. If an external field with frequency $\omega_p = \omega - g$ resonantly induces the single-photon transition $\ket{0,g} \rightarrow \ket{-,1}_\text{JC}$, the subsequent transitions $\ket{-,1}_\text{JC} \rightarrow \ket{\pm,2}_\text{JC}$ do not occur for $g$ strong enough, since their transition frequencies do not match $\omega_p$, leading to a single-photon blockade.}
    \label{j_c_ladder}
\end{figure}

Consider a two-level system, with transition frequency $\omega$, resonantly coupled to a cavity mode through a JC interaction. This system, when driven by an external field with frequency $\omega_p$ and modulated amplitude $\varepsilon(t)$, is described, in an interaction picture rotating at frequency $\omega_p$, by ($\hbar = 1$)
\begin{align} \label{HJC}
    H_\text{JC} &=  \Delta a^\dagger a  + \Delta\frac{\sigma_z}{2} + g(a\sigma_{+} + a^{\dagger }\sigma_{-}) \nonumber \\ 
    &+ \varepsilon(t)[(1-\lambda) (a\dag + a) + \lambda(\sigma_{+} + \sigma_{-})],
\end{align}
with $\Delta = \omega-\omega_p$ and $\lambda = \{0,1\}$. When $\lambda = 0$, the external field drives the cavity, while it drives the atom when $\lambda = 1$.

The eigenstates of the JC model are \cite{walls}
\begin{align}
        &\ket{g,0}, {\text{for} ~n = 0}\\
        &\ket{\pm,n}_\text{JC}=\frac{1}{\sqrt{2}}(\ket{g,n}\pm \ket{e,n-1}),\: \text{for}\: n\geq1,
    \end{align}
with respective eigenenergies
\begin{align}
    E_{g,{0}}^\text{JC} &= -\omega/2 ,\\
        E_{\pm,n}^\text{JC}&=\left(n-1/2\right)\omega \pm g\sqrt{n}\,.
\end{align}

Assuming the system is initially in the ground state $\ket{g,0}$, the driving field can resonantly excite it to its first dressed states $(\ket{g,0} \to \ket{\pm,1}_\text{JC})$ when tuned to $\omega_p = \omega \pm g$ ($\Delta = \pm g$). However, if $g$ is sufficiently strong, the driving field cannot populate the second dressed states and so on, as the transition frequencies from the first to the second dressed states are not resonant to $\omega_p$. \textbf{Figure} \ref{j_c_ladder} illustrates the eigenstates of the JC system and the transitions induced by a driving field with $\omega_p = \omega - g$. Here, the analysis is equivalent if the system is driven by either the cavity ($\lambda = 0$) or the atom ($\lambda = 1$). 




\begin{acknowledgments}
Luiz O. R. Solak would like to acknowledge Mr.~Miguel Cece for the fruitful discussions in the earlier stage of this work. This study was financed, in part, by the São Paulo Research Foundation (FAPESP), Brazil, Process Numbers 2022/00209-6 and 2023/09215-1, by Coordenação de Aperfeiçoamento de Pessoal de Nível Superior - Brasil (CAPES) - Finance Code 001, by CAPES-COFECUB (CAPES, Grant No.~88887.711967/2022-00), and by the Brazilian National Council for Scientific and Technological Development -- CNPq, Grants No.~405712/2023-5 and No.~311612/2021-0.
\end{acknowledgments}


\bibliography{apssamp.bib}

\end{document}